\documentclass[acmlarge,,anonymous=false]{acmart}

\AtBeginDocument{%
  \providecommand\BibTeX{{%
    \normalfont B\kern-0.5em{\scshape i\kern-0.25em b}\kern-0.8em\TeX}}}

\setcopyright{rightsretained}
\copyrightyear{2020}
\acmYear{2020}

\acmBooktitle{}
\acmPrice{}
\acmISBN{}

\usepackage{multirow}
\usepackage{booktabs}
\usepackage{rotating}
\usepackage{amsthm}
\usepackage{subcaption}
\usepackage[ruled,vlined]{algorithm2e}
\usepackage{enumitem}
\usepackage[title]{appendix}

\begin{document}

\title{PAR: Personal Activity Radius Camera View for Contextual Sensing}

\author{Jessica Maria Echterhoff}
\email{jechterh@eng.ucsd.edu}
\affiliation{%
  \institution{University of California, San Diego}
}
\author{Edward J. Wang}
\email{ejaywang@eng.ucsd.edu}
\affiliation{%
    \institution{University of California, San Diego}
}
\orcid{}
\renewcommand{\shortauthors}{Echterhoff and Wang}

\begin{abstract}
  Contextual sensing using wearable cameras has seen a variety of different camera angles proposed to capture a wide gamut of different visual scenes. In this paper, we propose a new camera view that aims to capture the same visual information as many of the camera positions and orientations combined from a single camera view point. The camera, mounted on the corner of a glasses frame is pointing downwards towards the floor, a field-of-view we named Personal Activity Radius (PAR). The PAR field-of-view captures the visual information around a wearer`s personal bubble, including items they interact with, their body motion, their surrounding environment, etc. In our evaluation, we tested the PAR view`s interpretability by human labelers in two different activity tracking scenarios: food related behaviors and exercise tracking. Human labelers achieved an overall high level of precision in identifying body motions in exercise tracking of 91\% precision and eating/drinking motions at 96\% precision. Item interaction identification reached a precision of 86\% precision for labeling grocery categories. We show a high level on the device setup and contextual views we were able to capture with the device. We see that the camera wide angle captures different activities such as driving, shopping, gym exercises, walking and eating and can observe the specific interaction item of the user as well as the immediate contextual surrounding.
\end{abstract}


\maketitle
\begin{figure}[h]
  \centering
  \includegraphics[width=\linewidth]{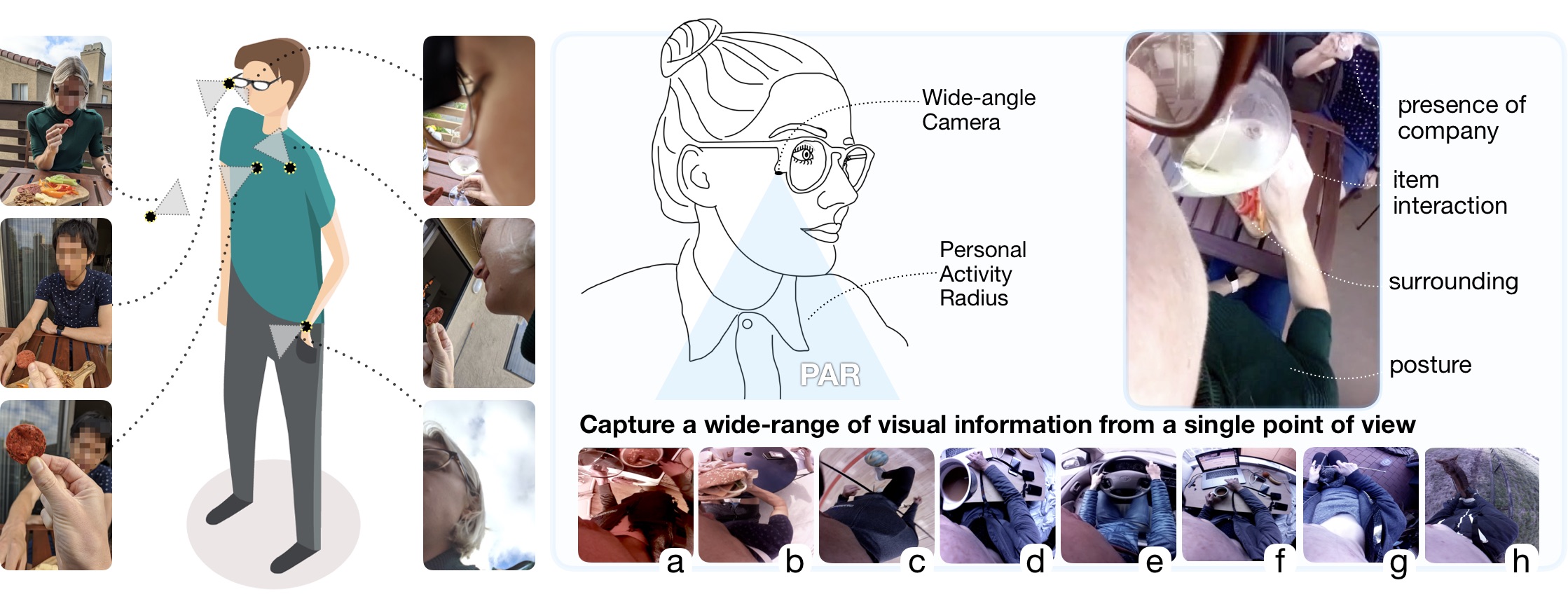}
  \setlength{\abovecaptionskip}{-15pt}
  \setlength{\belowcaptionskip}{-10pt}
  \caption{Left: Examples of what a scene would look like in the various camera views proposed in prior work. Right Top: Proposed camera setup showing a downwards facing capturing angle of a camera within our hardware setup. Right Bottom: PAR view of (a) cooking, (b) burger eating, (c) kicking a ball, (d) drinking coffee, (e) driving, (f) working on computer, (g) knitting, (h) riding a horse.}
  \Description{}
  \label{fig:intro}
\end{figure}

\section{Introduction \& Background}

Measuring a person`s behavior through understanding their context and activity is central to enabling a variety of ubiquitous computing applications for mobile and wearable systems \citep{abowd1999towards}. In the frame of wearable health monitoring, tracking and understanding a person`s behavior is needed to complement physiological measures to truly understand the dynamical nature of the person`s body response in order to fight diseases, improve living quality, and provide awareness for habits in someone`s life \citep{4911602, kalantarian2014wearable, souza2017wellness}. In the frame of rich and seamless interactivity of mobile computers, knowing the context can help determine what information to serve, when to serve it, and how best to adapt it based on the user`s needs now and in the future \citep{fischer2011investigating}. Things such as time, location, mobility information, and gross body movement are largely solved problems \citep{zhan2012activity, bulling2014tutorial}. However, fine grained semantics like what surrounds the person at a specific location, what are the different activities the person engaged in at said location, the quality of the person`s actual movement and more are particularly difficult to capture as the variety of what a person does over time is widely varying and what we want to measure is often application dependent. 



Wearable cameras have been shown to enable tracking a wide-range of human activities and behaviors \citep{starner1998visual, cheng2001personal}. With the camera pointed at a subject`s face, cameras can track the mouth for potential food or water intake or habitual activities like nail biting \citep{Bedri2020Fit, zhang2019necksense}. Pointed out at the chest and shoulder level, hand manipulations can be seen in front of the body and the item of interest often gets captured \citep{healey1998startlecam, ren2009egocentric, lee2015predicting,castle2010combining, mayol2004interaction}. At the waist and lower body level, cameras can observe a person`s posture and ambulatory movements, surrounding and walking conditions \citep{mariakakis2016watchudrive, yamazoe2005body}. Cameras attached to a human`s head capture full body motions like squatting, bending, push-ups, and other motions where the head moves in space, which are captured as an optical flow of the environment in conjunction with the body posture change \citep{zhan2012activity}. 

Wearable cameras are indeed a very flexible tool for capturing a whole gamut of information, but suffer from a major limitation. Each wearable camera is limited to where their field-of-view is pointed. If we need to move the camera for each application, it requires an active interaction from the wearer to move the camera for each use case. This makes unobserved background processing of many activity tracking applications almost impossible. If instead, a single physical orientation can capture the majority of visual information needed for the many applications proposed thus far, we can imagine developing a wide-range of software-enabled "virtual sensors". We use a camera orientation on a glasses form-factor where a wide-angle camera is positioned at the edge of the frame with the field of view pointed downwards into the wearer`s personal bubble. We named this field of view the Personal Activity Radius (PAR). The PAR field-of-view is a subset of the Peripersonal Space concept used in neuroscience~\citep{teramoto2018behavioral} and constrains the visual scene to a view on the wearer from above, specifically on everything below the user`s nose and down to the ground with a radius of about an arm`s length. Figure \ref{fig:intro} shows a sampling of the various scenes captured by our prototype PAR camera glasses. The PAR field-of-view encapsulates a wide-range of visual contexts of interest to the wearable camera community and activity/behavior tracking research topics. 

In this paper, we focus on assessing the visual data captured by a PAR sensor for a variety of contextual applications. To assess the potential interpretability of videos captured in the PAR field-of-view, we focused on how well these videos can be human labeled for future computational method developments or for journaling applications. To do so, we conducted a user study around two classic activity recognition scenarios: Food related behaviors and exercise tracking. In our evaluation, we recorded six participants performing four grocery shopping sessions, 15 eating sessions, and seven gym workout sessions. Reviewing videos captured in the PAR view, human labelers were able to correctly identify grocery categories with 86\% precision (w/ 7-classes), eating/drinking gestures with 96\% precision (w/ 3-classes), context of eating with 89\% precision (w/ 4-classes), exercise equipment at 94\% precision (w/ 4-classes), and exercise moves at 91\% precision (w/ 17-classes). Exercise repetition counting is also encapsulated in the PAR view, with counting accuracy at 89\%.


\section{Related Work}
\subsection{Discrete Camera Angles in Wearable Computing}
Wearable cameras have been explored for the last two decades and play an important role in recognition tasks. Due to the ease of human interpret-ability, they are a widespread tool and often first measure to capture human context and activities. The general usability and usefulness of camera glasses pointing forward has been explored and evaluated in many academic research papers \citep{bipat2019analyzing, ye2015detecting, kanade2012first}. Different camera setups have been explored and their resolution and capturing mechanisms have been refined over the years. Figure \ref{fig:intro} (left) shows existing and previously evaluated camera angles and locations on the user`s body which have been used in modern research for activity and object recognition, human behavior analysis and contextual evaluation.

A first person field of view for object detection has been explored by Lee and Grauman to provide a storyboard clustering of the wearer`s day including an estimate of important objects and people \citep{lee2015predicting}. A similar setup exists in form of front facing cameras on the chest of a human`s body. These were used for capturing memorable events \citep{healey1998startlecam}.
A head-mounted camera has been explored by Yamazoe et al. to estimate head pose, human activities and aims to provide possibilities of a digital diary with their device setup \citep{yamazoe2007body}. Kitani et al. explored head-/helmet-mounted cameras for sports activity clips to learn ego-action categories and cluster video content \citep{kitani2011fast}.
Drinking, walking, as well as activities such as going upstairs and downstairs have previously been captured by a front facing camera attached to a user`s glasses and evaluated using machine learning methods. Those were proved to achieve an accuracy of 68.5\%-82.1\% \citep{zhan2012activity}.
Different researchers used a shoulder mounted camera to automatically detect different objects in day to day scenarios \citep{castle2010combining, mayol2004interaction}.

Similar setups using cameras were used to track and assess diets \citep{o2013using, wang2002validity}. Smartphones have been utilized to track diets and count calories with integrated smartphone cameras by taking pictures of the food and its residues \citep{kong2012dietcam}.
Researchers have used chest mounted cameras to analyse different interactions with objects \citep{ren2009egocentric}. In order to save bandwidth and processing power, researchers used smartphones in a regular shirt chest pocket. In these setups, cameras and underlying sensors filter ques for when to take images~\citep{han2014glimpsedata}. Another field of view was proposed with a body camera pointing upwards to capture the subjects face and upper body from below and the additional use of other sensors for activity recognition \citep{yamazoe2005body}. A camera pointing upwards for ground truth data collection was used in \citep{zhang2019necksense}. Similarly, researchers previously experimented with a camera attached to a subjects wrists for differentiating drivers and passengers in cars \citep{mariakakis2016watchudrive}.

Starner et al. used a two camera setup mounted onto a hat capturing the body and hands with a downwards facing camera and a forward facing camera capturing the user`s field of view for hand pose and contextual awareness \citep{starner1998visual}. Another use case for a downwards facing camera, implemented by a forward pointing camera device on glasses was proposed by Cheng et al. by introducing dance step instructions when the subject was looking downwards \citep{cheng2001personal}. A camera capturing a user`s mouth for food intake recognition has been used by Bedri et al. They used a sidewards angeled camera mounted on a pair of glasses and included multimodal sensing to get a broader picture in dietary monitoring and determine operation of a camera based on sensor queues \citep{Bedri2020Fit}.

The position of cameras in a third person view has been used for spy cameras \citep{fiore2008multi} or simultaneous exercise recognition in gyms \citep{khurana2018gymcam}. This approach has its own, slightly different use cases than the wearable approach majorly discussed and explored in our work.

\subsection{Interaction with Items}
We all interact with hundreds of items every day visually and physically. The field of computer vision studied automatic object recognition and detection from videos and images \cite{redmon2016you, papageorgiou1998general, szegedy2013deep, girshick2015fast}. Market research is interested to see which items we select to interact with out of the almost unlimited selection that is offered to us. In the virtual world, we track the web surfing behavior to determine interest into a specific product\citep{koufaris2001consumer}. But in the real world, the extend of item interaction is hard to determine and evaluate without context. Typically, visual item interaction has been approached by gaze tracking and forward facing cameras \citep{smith2013gaze, lee2015predicting}. But different sensors, for example magneto-inductive sensors, self-powered radio tags and non the less cameras attached to specific devices can give insights into stationary human-device interaction \citep{wang2015magnifisense, zhang2019sozu}. 

\subsection{Scene Analysis}
Scene analysis is an integrated part of context aware ubiquitous computing \citep{abowd1999towards}. Several sub-domains have  explored multiple sensors setups to achieve an ubiquitous caption of a user`s context. Cameras as well as multimodal sensing techniques with a variety of camera angles capture human behavior have previously been explored for recognition and contextual analysis. 
Scene analysis has been explored to enrich activity recognition tasks with additional information. It was shown to be used as a technique to facilitate distinction between activities which share common properties and make use of dependencies of contextual happenings and their implications on recognition tasks \citep{zhu2012context}. Smartphones and smartwatches and their integrated sensors have been used to explain diverse human context \citep{vaizman2017recognizing, vaizman2018context,echterhoff2018gait2}. Wide area sensing methods have been used to gather information around a scene \citep{laput2019exploring, khurana2018gymcam}. Multimodal sensing techniques enabled the fusion of different sensors into convenient form factors \citep{Bedri2020Fit}. Glasses as one form factor have been used (among other use cases) for contextual sensing for alertness measurement using electrooculography sensors \citep{tag2019continuous}. 

\subsection{Body Pose \& Motion}
Determining how much time a user spent sitting, standing or moving around between different poses is beneficial for a granular estimation of activity density throughout a user`s day and gives also hints about habits and life circumstances of the user. Pose estimation has been an interesting research topic for computer vision, where most setups use deep neural networks to classify different poses in image sequences \citep{newell2016stacked, toshev2014deeppose}. Since poses are a static action it is almost impossible to capture them with regular on-body sensors. However, Laput and Harisson proposed a wrist worn device that senses a variety of hand activities and might infer poses from these \citep{laput2019sensing}.

Accelerometers, gyroscopes, acoustic sensors and cameras are very popular sensors for motion recognition \citep{ha2016convolutional, echterhoff2018gait}. These sensors have been explored both on wearable devices as well as in stationary placed sensor-to-object setups. For example, electromyography signals \citep{zhang2016diet}, proximity sensors \citep{chun2018detecting} or direct on-body sensors such as a microphone \citep{amft2008recognition} have been used to track eating behavior with jaw movements.

Repetition in human motion can be indicated in two different ways. The existence of multiple items can hint towards a repeated action, for example when picking up three apples from the grocery store. Repetition can also occur with temporal spatial advancing motion or optical flow. Thus, item quantity can also be observed by tracking repetitions of hand motions. This has also been tackled from the computer vision side as well as with accelerometers and gyroscopes \citep{levy2015live,mortazavi2014determining}. Recognition of repetitions can therefore give indications on quantitative nutritional analysis of occasional or frequent eating or snacking behavior.

We also keep track of our physical fitness by tracking repetitions of the same activity, for example when we go to the gym. Repetition is an important factor to measure intensity and duration of an activity and has been explored using cameras and neural convolutional neural networks \citep{levy2015live, khurana2018gymcam}.

\section{Personal Activity Radius}
\subsection{PAR Field-of-View}
The PAR view can best be described as how it looks when you tilt your head down and look with one eye closed. Your mouth occupies the bottom left corner, the shoulder and torso filling the bottom. When your arm reaches out at torso level, it appears in the middle of the view. You can see objects that enter your personal bubble and interactions you make with your hands and feet can be seen. As you move, the floor moves beneath you. If you do a push up or sit up, you see your torso move with respect to the surrounding. Your surroundings such as what objects are around or if you are in the presence of company is also captured. What your eye sees in this perspective is captured by in the PAR field-of-view. 

\subsubsection{Interaction with Items}
The PAR view captures an image centralized around a person`s arm movements, which offers a special perspective on interactions with items. Those interactions frequently occur with the user`s hands. The PAR view can see different proximity levels of items respective to the user`s body and captures the duration and manner of hand to item interaction, for example when picking up an apple to tell if it`s quality is sufficient for a purchasing action. Similar actions occur when eating or drinking, where we see a hand-to-mouth action with a respective item, such as a fork, knife or drinking bottle. Thus, the PAR view also captures near-to-mouth activity due to it`s slight angle. This angle captures the mouth and lower nose and therefore also captures when a person touches his or her face, for example with a napkin, but does not intake any food. The view can also see an interaction of other limbs with objects, such as the interaction of the foot with a ball and can see the pattern and duration of ball contact of a player. 
	
\subsubsection{Scene Analysis}
The PAR view captures the surrounding environment of the user, which is defined by the scene. Changes in scene such as the physical location and the surrounding objects in different proximity to the user can be observed. A person can be in a scene defined by a large objects that surround the wearer, such as a car, a workstation, or even something like the back of a horse. 
\subsubsection{Body Pose \& Motion}
Body motion is captured via changes of body parts towards the sensor. Those changes occur through spatial deviations of body parts over time. When a person for example performs a leg lift, operates weight lifting machines, runs or rides a bike, the PAR view shows a change in distance of the legs and feet to the sensor. Curls, lifting and bench press exercises as well as grabbing items are captured with the up and down or back and forth arm movement, where the hands appear in different scales depending on proximity to the sensor. A similar change in proximity is captured when the PAR sensor itself is moved. Those movements happen when performing sit ups, where the sensor itself changes its position in a temporal manner towards the knees and back to the floor. Push ups, planks or mountain climbers show the movement of the whole view towards the floor or to the left and right, which indicates head movements. The PAR view also captures motions combining the spatial deviation of the sensor and body part movements. When a person runs, uses a rowing machine or performs jumping jacks, the view itself moves (up and down as well as back and forth respectively) as well as body parts move towards the sensor, for example arm or leg movements. Full body motion are seen through the environment moving because the camera is mounted to the head. 

Repetition within the PAR view is captured through the same body motion or item interaction appearing multiple times within a short time frame. For example, if a person is picking up three apples, the repeated consecutive arm motion reaching for those apples is seen over time within the view. Also, the number of apples is shown to support the number of body motions. When there is no item as a repetition count support, the PAR view infers repetition through the change of view and proximity towards a reference. For example, when a person is doing five push ups, we can see the view changing its position repetitively to the floor, which creates a change of perspective five times in a row.

\subsection{PAR Wearable Camera Glasses}
\begin{figure}[ht]
  \centering
  \includegraphics[width=\linewidth]{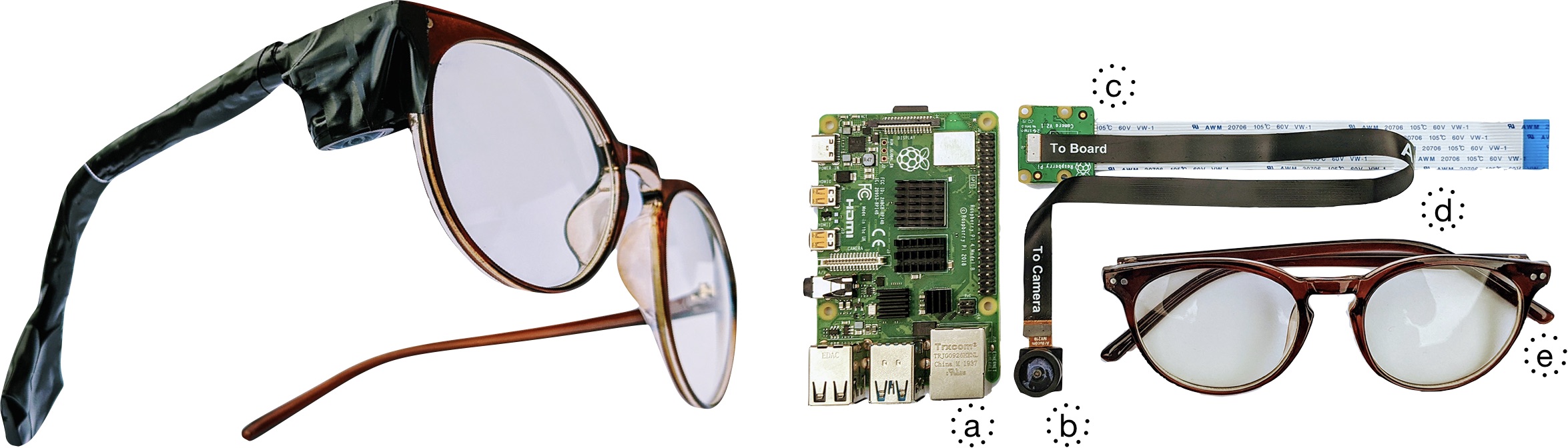}
  \setlength{\belowcaptionskip}{-10pt}
  \caption{Left: Proof-of-concept PAR Camera Glasses. Right: (a) RPi 4, (b) IMX219 170\textdegree Camera, (c) RPi Camera Module V2.1, (d) ArduCam 300mm Extension Cable, and (e) glasses frame}
  \Description{}
  \label{fig:prototype}
\end{figure}

We constructed a proof-of-concept wearable camera setup, shown in Figure \ref{fig:prototype}, with an 8 megapixel camera (IMX219), including a 170\textdegree, M-12 lens mounted on a pair of glasses. The attached camera points downwards to capture the PAR of the user with the intent to not interfere with or directly capture other people within the scene. The camera lens is connected to a Raspberry Pi Camera Module V2.1 via an ArduCam 300mm extension cable. The cable is wrapped along the glasses arm and attached to the camera module which is attached to the end of the frame arm (behind the ear). The extension cable is wrapped to reduce crimps on the cable, which introduces signal drop-offs. Finally the camera module is connected to a Raspberry Pi 4 via a CSI cable. The processing component and the camera was powered using an off-the-shelf battery pack. Construction instructions and accompanying software will be made available via the following online repository (\textit{https://gitlab.com/ubicomp-lab/ubicompi}). All data captured with the PAR glasses including the respective labels is made available at \textit{https://gitlab.com/ubicomp-lab/par\_data}.

\section{Evaluation}
 In this paper, we evaluate the system by looking at a number of prototypical activity, context recognition and tracking scenarios. In each scenario, we designed a set of use cases to capture how the visual data looks in the PAR view. The evaluation focuses on examining the label-ability of the visual data by crowd-sourced human labelers. The label-ability of the visual data is extremely important as further development of automated systems, using PAR as the source of visual data, would require labeling of the scene based on their application. If the visual information captured by PAR is too difficult to interpret by human labelers, it would mean that videos would need to be labeled at the time of capture to distill the data collector`s interpretation of the scene, losing one of the biggest strength of vision-based analysis. The different scenarios help to reveal aspects around the PAR view in regards to tasks commonly desired for visual computing such as object identification \citep{redmon2016you}, body motion \citep{khurana2018gymcam} and posture analysis \citep{toshev2014deeppose}, and the visibility and interpretability of the surrounding \citep{furnari2015recognizing}. We suspect that end-applications would dictate the type of algorithm ultimately employed, but the label-ability of different components of the scene that we examine here provides a generalizable understanding of how our solution can provide machine learnable visual data. All procedures are approved under our institution`s IRB. 
 
\subsection{Scenario 1: Food Related Behavior}
 \begin{figure}[!ht]
  \centering
  \includegraphics[width=\linewidth]{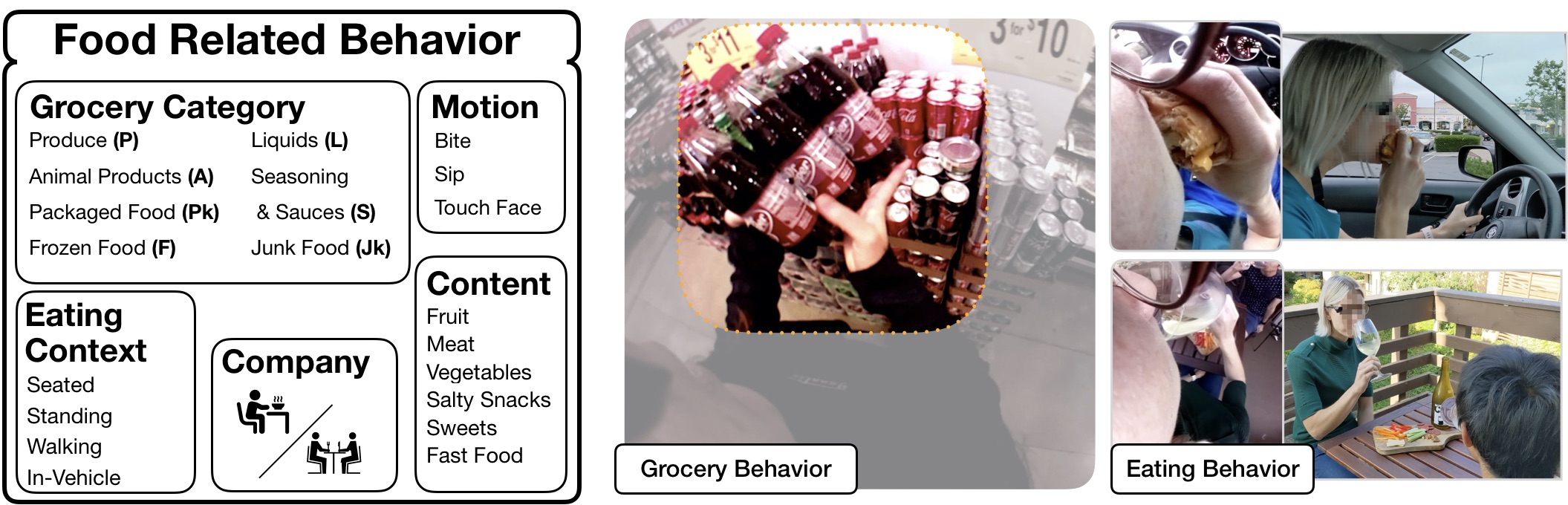}
  \setlength{\abovecaptionskip}{-5pt}
  \setlength{\belowcaptionskip}{-10pt}
  \caption{We evaluated human behavior by observing interactions with food items in a grocery store. We additionally evaluated eating motion within specified contexts, the respective food content and if the wearer was eating alone or with company.}
  \Description{}
  \label{fig:shopping_eval}
\end{figure}

We designed a shopping behavior scenario, where we observe a subject during a regular grocery shopping trip. We defined a grocery item list capturing the following categories: produce, liquids, animal products, frozen food, packaged food, seasoning and sauces as well as junk food. The shopping list was designed to represent a variety of items from a western diet. To diversify the duration of interaction, subjects were cued to either interact with the item quickly (<1 second) or for a longer duration (inspect the item for a self-determined 2-5 seconds). Each of the 3 subjects visited one grocery store, with one of the participants visiting two grocery stores, providing data for 4 different grocery stores. In total, we collected 64 minutes and 57 seconds across the 8 different grocery categories. 

Within the eating activity scenario, our experiment is designed to observe a subject during a eating situation where they freely consume food item of their choice (i.e. ordered food, self-prepared food, or snack). We observed each person within a 5-minute eating interval and annotate each bite and sip. To add additional complexity to the data, we also collected instances of the subjects just bringing their hand to their face without eating or drinking. The general contents for food items were annotated, along with the time-stamp of each time the hand reaches towards the mouth. We additionally labeled the user`s context while eating, focused on if the person is seated, standing, walking, or commuting in-vehicle while eating. For one of the eating session, we conducted a shared dining environment of 4 people sitting around the table (one on either side of the wearer and another across the table). This session is used to to observe how well the PAR view can capture company of others during meals.



\medskip

\textbf{Analysis:} We performed four analyses on the labeling performance: Grocery Classification, Eating Motion Classification, Food Content Identification, and Eating Context Classification. Furthermore, we analyze the percentage of time the PAR view captures people sitting around a table during a meal.

\smallskip

 \noindent\underline{Grocery Classification}. The human labelers showed an average accuracy of 71\% for correctly labeling grocery shopping items across 7 categories. For purposes of labeling performance, the precision metric is particularly crucial since we want to be able to trust that the labels are indeed what they say they are. When looking at the precision, we see that the average precision is 86\%, with 5 classes having >=90\% precision. Packaged food is the only class that performed lower than 80\%, with a fairly low 47\%. This may be a case where the semantic of packaged food is unclear since technically most of the food items were indeed "in a package" so could have been misconstrued as "packaged food." In contrast, the recall is comparatively low, with an average of 70\%, and only 2 classes having >=90\% recall. A low recall may not be as big of an issue given that we included a None of the Above case, with most of the missed instances falling under the None class. That means that some categories people are less confident about, for example liquids and junk food are largely missed as None. Item level classification can be seen in Appendix A.

\begin{figure}[t!]
  \centering
  \includegraphics[width=\linewidth]{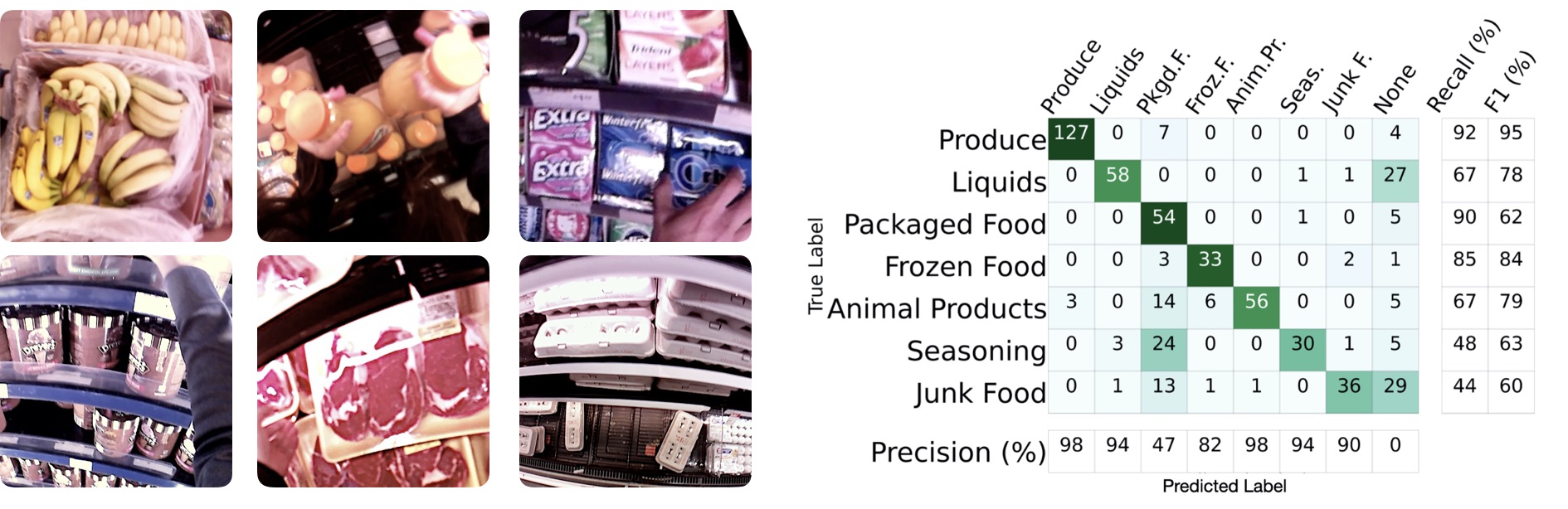}
  \setlength{\abovecaptionskip}{-10pt}
  \setlength{\belowcaptionskip}{-10pt}
  \caption{Left: Examples of grocery items captured in PAR. Right: Confusion matrix of grocery category classification for food items.}
  \Description{}
  \label{fig:shopping_analysis}
\end{figure}

\smallskip
 
\noindent\underline{Eating Motion Classification}. When asked to identify the hand to face motion, the labelers accurately distinguished between taking a bite, having a sip, and just touching the face with 94\% accuracy. All classes achieved at least 96\% precision. Face touching is the most misconstrued class with the lowest recall of 81\% and 96\% precision.  

\begin{figure}[b]
  \centering
  \includegraphics[width=\linewidth]{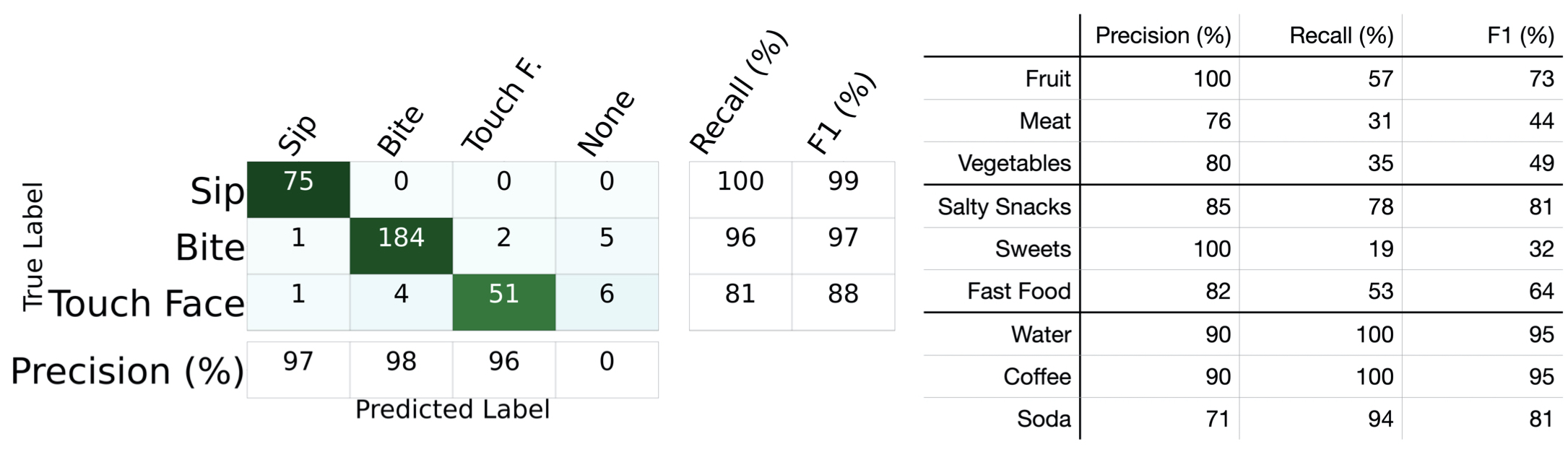}
  \setlength{\abovecaptionskip}{-10pt}
  \setlength{\belowcaptionskip}{-10pt}
  \caption{Left: Confusion matrix of labeling eating motion. Right: Labeling accuracy of food content in the scene}
  \Description{}
  \label{fig:eating_eval}
\end{figure}

\smallskip

\noindent\underline{Food Content Identification}. After the human labelers identified sips or bites, they are then asked to identify the content of the food visible in the scene. Because each food instance may contain more than one food content, they can select more than one category. As seen in Figure \ref{fig:eating_eval}, the average precision is much higher than average recall, 86\% compared to 63\%. For examples of salty snacks and sweets, most of our examples are food items that are quite small (e.g. nuts, gold fish crackers). When reviewing the videos, even the research team is only able to confidently identify what was being eaten by looking at the package or plate and not what was in the hand at the time of consumption. We speculate that this was likely the case for the crowd labelers given the fairly high precision score. This is an interesting advantage of the PAR view as compared to a more face focused visual view. PAR is able to provide simultaneously when food is being placed in the mouth, but also provide the scene of the table which often contains the plate or packaging, giving more context that can be used to disambiguate. We discuss this in more detail in the discussion section. 

\smallskip

\noindent\underline{Eating Context Classification}. While observing the eating context scenario, we focused on the postural-locational context of eating. We evaluated the scenario in terms of if the subjects are seated, standing, walking, or eating in a vehicle while being a passenger or a driver. These contexts were classified correctly by the labelers with an accuracy of 88\%. We can see some confusion between walking and standing.

\smallskip
 \begin{figure}[t!]
  \centering
  \includegraphics[width=\linewidth]{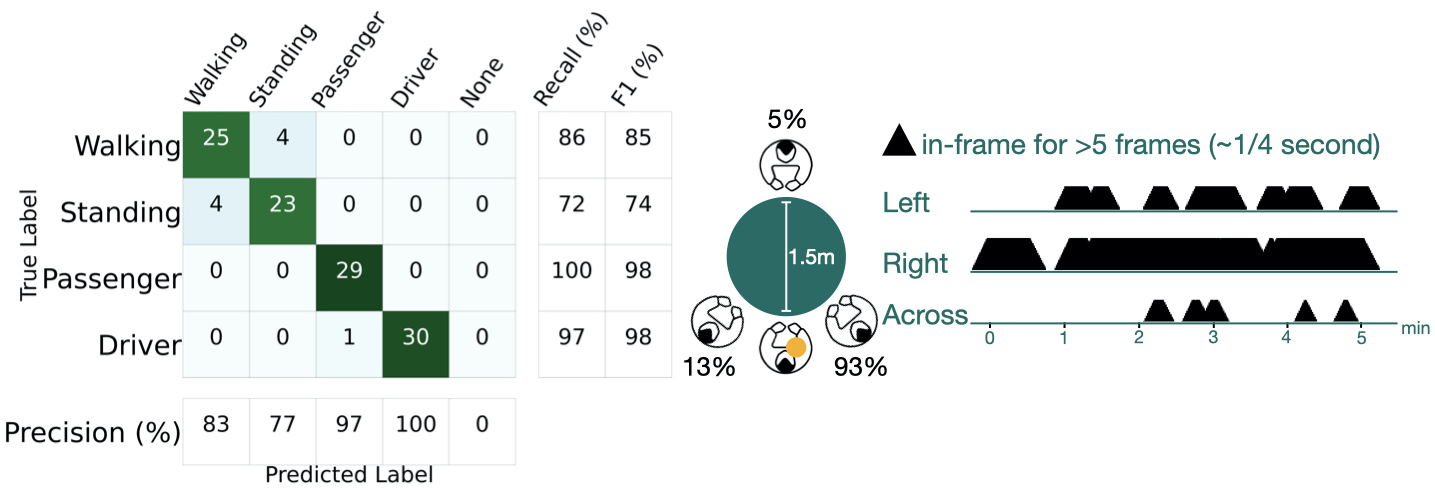}
  \setlength{\abovecaptionskip}{-10pt}
  \setlength{\belowcaptionskip}{-10pt}
  \caption{Left: Confusion matrix for the eating context labeling. Right: Percentage of time when one or more full body parts of a person at a dining table are in frame captured by the PAR view.}
  \Description{}
  \label{fig:eating_context}
\end{figure}
 \noindent\underline{Presence of Company}. For examining the ability of the PAR view at capturing the presence of company, we looked at how often a person would be in view if they were seated at different locations around a table where the wearer is eating. The camera is located on the right corner of the glasses, so the left side view is more obstructed. Figure \ref{fig:eating_context}b shows the position of the different people. The person to the right is most frequently in the frame, with 93\% of the total 5 minute eating session. The person to the left is less frequently in the frame at 13\%. The person across the subject is captured 5\% of the time. The person to the left and across the table appear in the frame when the wearer moves their head between bites or redirects their attention, say during conversation. Hands can also show up in the frame. The table we chose is 1.5 meters in diameter. For a smaller dining table, we imagine the person sitting across would be in the frame more frequently. However, the purpose of knowing if there is company during eating is not to track the behavior of others but rather just to determine the presence of others. Even at the lowest frequency of identification of the opposite side of the camera (in this case the left side), given in a 5 minute eating session, the person is seen at least a few times.

\subsection{Scenario 2: Exercise Tracking}
 \begin{figure}[h!]
  \centering
  \includegraphics[width=\linewidth]{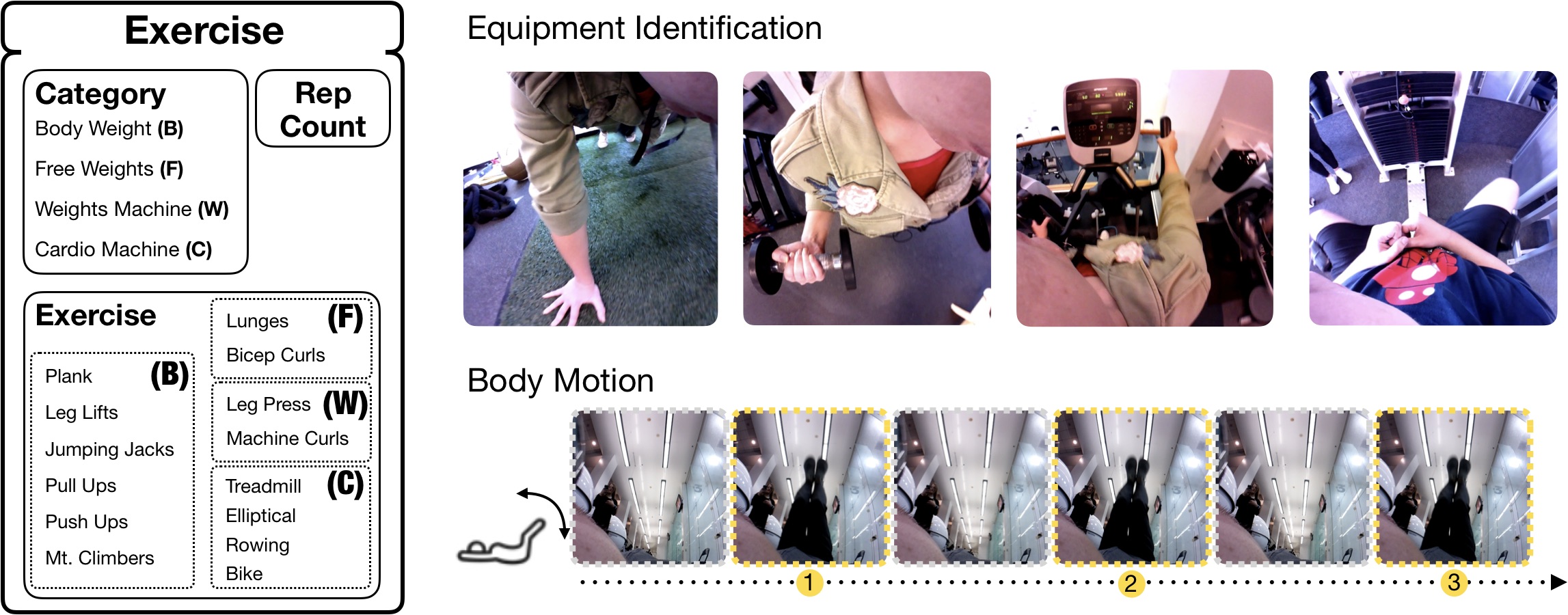}
  \setlength{\abovecaptionskip}{-5pt}
  \setlength{\belowcaptionskip}{-10pt}
  \caption{In the exercise scenario, we asked subjects to exercise using different equipment, and perform a number of different exercise moves with each kind of equipment. Right: Examples of different exercise equipment such as body weight (push up), free weights (dumbbell), cardio machine (elliptical), and weights machine (leg machine). Participants also performed exercises with different parts of their body such as leg lifts.}
  \Description{}
  \label{fig:exercise_eval}
\end{figure}

In the exercise activity scenario, the experiment is designed to observe a subject doing a variety of exercises in an indoor gym environment. We are interested in seeing how well PAR captures the exercise being performed both in terms of the different equipment and also body movement and movement repetition. We observed typical movements trackable by accelerometer based systems, which typically include hand movement or full body motion, such as arm curls, jumping jacks, treadmills. We also included more stationary movements for which accelerometers don`t typically perform well with, such as push ups, leg lifts, and biking. 

\medskip

\textbf{Data Collection:} We selected activities within four different categories of equipment: body weight, free weights, weight machines, and cardio machines. For each equipment, a variety of different exercises were performed. Activities that are repetition based (e.g. push-ups, bicep curls) are performed for ten repetitions. Activities on cardio machines are performed for one minute. Exercises were performed in a randomized sequence. The exercise assignment is announced to the subject in the same way as the shopping activity. In total, we collected data from 6 people in 4 different gyms. 



\medskip

\textbf{Analysis:} We performed three analyses on the labeling performance: Exercise Equipment Classification, Exercise Move Classification, and Repetition Counting. 

 \begin{figure}[b]
  \centering
  \includegraphics[width=\linewidth]{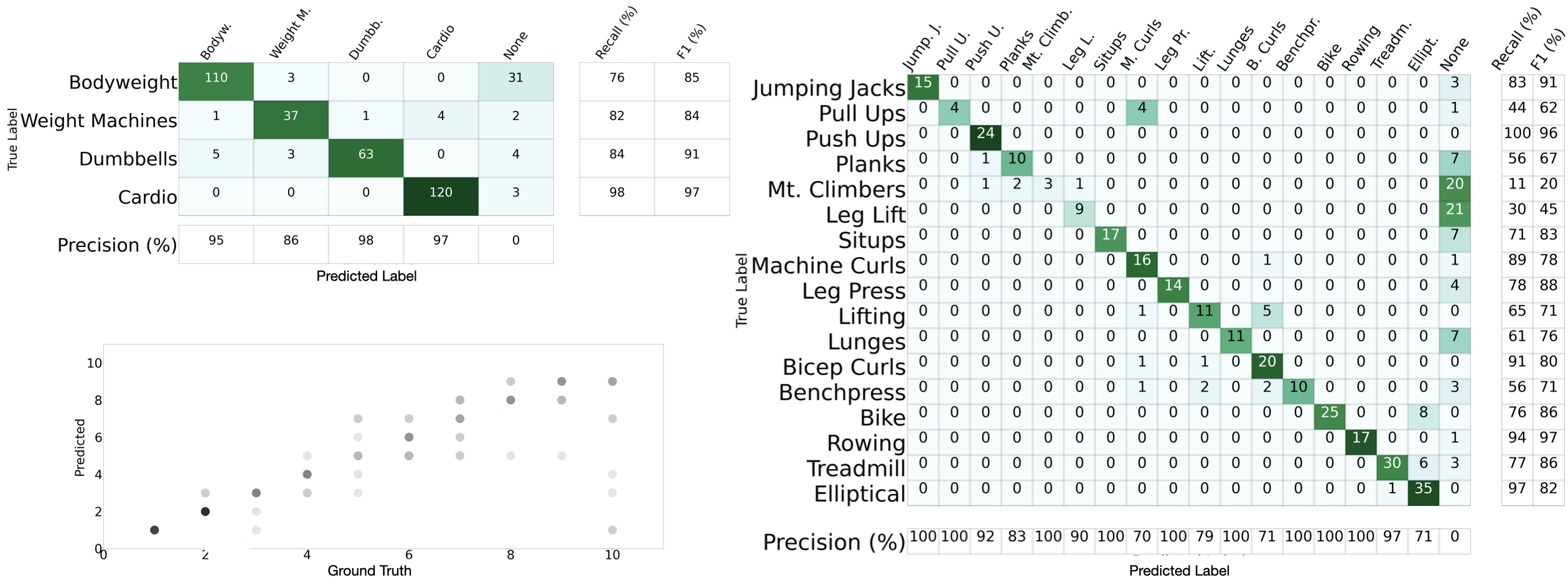}
  \caption{Left Top: Confusion matrix of equipment category labeling. Right: Confusion matrix of exercise moves labeling. Left Bottom: Correlation plot of repetition counting.}
  \Description{}
  \label{fig:exercise_analysis}
\end{figure}

\smallskip

 \noindent\underline{Equipment classification}. When asked to identify the exercise based on the equipment used, the labelers accurately labeled exercise by the equipment used 85\% of the time. Just like in the shopping analysis, precision stands as the more crucial metric. The average precision is 94\%, with only one class (weights machine) lower than 95\% precision. For the most part, most of the confusion came down to labeling as None of the Above, leading to a comparatively low recall of 85\%. 

\smallskip

\noindent\underline{Exercise Move Classification}. When asked to identify what the actual exercise move is, the average labeling accuracy drops to 70\% across the the 17 exercise moves. Similar to the shopping item classification, this decrease in accuracy seems mostly attributed to a reduction in recall through labeling items as None of the Above rather than inter-class mislabeling. The average recall dropped to 70\%, with 67\% attributed to labeling as None. In comparison, the precision stayed high at 91\%, with 12 out of 17 classes with at least 90\%. 

Looking closer at the confusions, some of the lowest recall activities included pull ups (44\%), mountain climbers (11\%), and leg lifts (30\%) were all body weight exercises. When the investigation team watched the videos ourselves, we also recognize that these videos are particularly difficult to interpret. The pull-ups were mostly mislabeled as an arm machine. This likely occurred with the way our labeling occurs in an hierarchical way where the labeler first chooses the equipment and then the specific exercise move. For mountain climbers and leg lifts, the actual activity is happening at the leg. For mountain climbers, the leg movement is only seen if the person is looking straight down or back down at the leg. Leg lifts, as seen in Figure \ref{fig:exercise_eval}, seem to be a difficult to interpret scene where the leg moves to and away from the body. Leg lifts mostly got labeled as None of the Above. Labelers were not shown any examples of what an exercise looks like in the PAR view. This leads to some of these less obvious body movements being not understandable without prior knowledge. 

What is particularly promising about these results is that the performance remained relatively high for the wide variety of exercise types that was tested. As seen in the variety of classes, subjects performed exercises that have repetitions, but also exercises that are stationary like the body plank. Many exercises where an accelerometer based system would struggle, such as push-ups and biking, the PAR view captures them clearly. We imagine that a sensor fusion approach for motion measurements augmented with the PAR view for scene clarification would be able to capture all exercise movements nearly perfectly. 

\smallskip

 \noindent\underline{Reptition counting}. Subjects were instructed to perform exercises that are repetition based for 10 repetitions. Labelers, were shown a partial segment of the 10 exercises containing a random number of repetitions. They counted the number of repetitions correctly with an accuracy of 89\%, a pearson correlation coefficient of $R = 0.81$ and $RMSE = 1.83$. 85\% of the total incorrect rep count instances were +/- 1 rep from the true count. Initially, we had anticipated counting results to be much closer to perfect given how visually obvious the repetitions of an exercise can be seen in our initial pilot study. Upon closer reviewing of exercise videos, we noted that when subjects naturally exercised, the wearer often turned their head in the middle of the exercise. This leads to a change of what part of the body is in the frame. For arm-based exercises, this means that the arm moves in and out of the frame, which likely led to miscounting. 

\section{Discussion \& Conclusion}
\subsection{Limitations}

\subsubsection{Privacy}
Wearable cameras typically face the issue of capturing identifiable information of people other than the user him or herself. In addition, seeing the camera lens facing those surrounding people creates a sense of privacy infringement. In the PAR view, the majority of the time, the camera only ever captures the wearer themselves, who by wearing the PAR sensor is self-consenting to the video capture. In addition, with the camera pointed down and situated in the inner corner of the glasses frame, the fact that the user is wearing a camera is not obvious. Additional industrial designs could easily mask the fact that there exists a camera at all, especially with a smaller custom lens. However, it is still possible that our device will capture faces of different people surrounding the main user. This happens when performing exercising tasks lying down or bending downwards. 

When considering the use of the PAR view for all-day monitoring, there is the potential of capturing private moments during the day (paying for a purchase with a credit card, visiting the bathroom, etc.). These are similar challenges faced by other vision-based activity tracking. A variety of solutions can be used to address this. On-device edge computation, where the actual image data never leaves the processing unit of the wearable device is one possible solution. Here, a pre-defined set of interpretations (such as amount of time a user has been reading a book) is used to distill from the visual information. In the case where the actual image is needed, such as for food journaling or un-structured analysis, a trigger based camera system can be employed. For example, as in the case of \citep{Bedri2020Fit}, the camera is triggered by a proximity sensor. These two ideas can also be combined where triggering first reduces the amount of visual data actually captured and an edge compute solution is then employed for detecting if an unwanted scene is captured and discarded before saving the image data. 

\subsubsection{Blind Spots}
As discussed in the related work, many different fields of view have been explored for camera`s attached to the user`s body. Our system can capture different fields of views but it is, as of now, is limited to the right side of the user`s face. We can imagine extending the field of view to increase the captured view to both sides of the body by adding a second camera. We found that a single camera was able to capture most of the information needed, but situations like scratching the face or an object on the left side of the body about an arm length away is temporarily out of view. However, because the head naturally moves around and faces objects of interest, it is likely that the head reorients towards something on the left if it becomes relevant. 

The forward facing view looking out into the surrounding, replicating where your eyes are naturally pointed and sees (the eye view), is underrepresented by the PAR view. We believe that the PAR view is more properly suited for use cases regarding activity recognition of the wearer themselves, whereas the eye view is more around potential points of interest, as approached in the eye tracking community \citep{smith2013gaze}. The eye view also has advantages for wearable camera photography. However, the eye view camera suffers from the issue of privacy infringement of bystanders as this view inherently captures the bystander`s face directly. In terms of user interest tracking, the PAR view is able to provide the scene close to the person. Although we did not use this in our evaluation, during our construction, we found that if the wide angle camera is positioned higher up on the frame with the camera above the eye and pointed inwards and forward slightly, it is possible to capture the eye and the forward scene of the wearer. Because the camera has a 170 degree field of view, by only tilting the camera by about 10-15\%, we can capture the forward scene at eye level without losing the effect of the PAR capturing the rest of the body, with a little loss of information behind the wearer.

\subsection{PAR as an Applicable View for Wearable Cameras}
In this paper, we show that the PAR camera view is capable of capturing a wide range of visual information ranging from item interaction, body motion, pose, surrounding environment and situation. Thus, contextual properties are successfully captured by our PAR field of view. We show that the data generated with this specific view is interpret-able by human labelers, and subsequently that it is useful for future automated classification development. We imagine that using the PAR view in an orchestrated manner with other sensors, we can increase performance of multiple classification tasks and give valuable contextual information. For example, in the frame of eating detection, a lower power sensor like a proximity sensor could be used to trigger the camera to perform a scene analysis \citep{Bedri2020Fit}. For exercise, a location-based service or motion level tracking is representative of exercise before triggering the camera for more detailed scene analysis. The PAR view can thus be a powerful add-on for sensor setups in need of  visual data for contextual disambiguation for human activity, behavior, and context.

\bibliographystyle{ACM-Reference-Format}
\bibliography{sample-manuscript}


\begin{thebibliography}{55}


\ifx \showCODEN    \undefined \def \showCODEN     #1{\unskip}     \fi
\ifx \showDOI      \undefined \def \showDOI       #1{#1}\fi
\ifx \showISBNx    \undefined \def \showISBNx     #1{\unskip}     \fi
\ifx \showISBNxiii \undefined \def \showISBNxiii  #1{\unskip}     \fi
\ifx \showISSN     \undefined \def \showISSN      #1{\unskip}     \fi
\ifx \showLCCN     \undefined \def \showLCCN      #1{\unskip}     \fi
\ifx \shownote     \undefined \def \shownote      #1{#1}          \fi
\ifx \showarticletitle \undefined \def \showarticletitle #1{#1}   \fi
\ifx \showURL      \undefined \def \showURL       {\relax}        \fi
\providecommand\bibfield[2]{#2}
\providecommand\bibinfo[2]{#2}
\providecommand\natexlab[1]{#1}
\providecommand\showeprint[2][]{arXiv:#2}

\bibitem[\protect\citeauthoryear{Abowd, Dey, Brown, Davies, Smith, and
  Steggles}{Abowd et~al\mbox{.}}{1999}]%
        {abowd1999towards}
\bibfield{author}{\bibinfo{person}{Gregory~D Abowd}, \bibinfo{person}{Anind~K
  Dey}, \bibinfo{person}{Peter~J Brown}, \bibinfo{person}{Nigel Davies},
  \bibinfo{person}{Mark Smith}, {and} \bibinfo{person}{Pete Steggles}.}
  \bibinfo{year}{1999}\natexlab{}.
\newblock \showarticletitle{Towards a better understanding of context and
  context-awareness}. In \bibinfo{booktitle}{\emph{International symposium on
  handheld and ubiquitous computing}}. Springer, \bibinfo{pages}{304--307}.
\newblock


\bibitem[\protect\citeauthoryear{Amft and Tr{\"o}ster}{Amft and
  Tr{\"o}ster}{2008}]%
        {amft2008recognition}
\bibfield{author}{\bibinfo{person}{Oliver Amft} {and} \bibinfo{person}{Gerhard
  Tr{\"o}ster}.} \bibinfo{year}{2008}\natexlab{}.
\newblock \showarticletitle{Recognition of dietary activity events using
  on-body sensors}.
\newblock \bibinfo{journal}{\emph{Artificial intelligence in medicine}}
  \bibinfo{volume}{42}, \bibinfo{number}{2} (\bibinfo{year}{2008}),
  \bibinfo{pages}{121--136}.
\newblock


\bibitem[\protect\citeauthoryear{Bedri, Li, Khurana, Bhuwalka, and Goel}{Bedri
  et~al\mbox{.}}{2020}]%
        {Bedri2020Fit}
\bibfield{author}{\bibinfo{person}{Abdelkareem Bedri}, \bibinfo{person}{Diana
  Li}, \bibinfo{person}{Rushil Khurana}, \bibinfo{person}{Kunal Bhuwalka},
  {and} \bibinfo{person}{Mayank Goel}.} \bibinfo{year}{2020}\natexlab{}.
\newblock \showarticletitle{FitByte: Automatic Diet Monitoring in Unconstrained
  Situations Using Multimodal Sensing on Eyeglasses}. In
  \bibinfo{booktitle}{\emph{Proceedings of the 2020 CHI Conference on Human
  Factors in Computing Systems}}. \bibinfo{publisher}{Association for Computing
  Machinery}, \bibinfo{address}{New York, NY, USA}, \bibinfo{pages}{1–12}.
\newblock
\showISBNx{9781450367080}


\bibitem[\protect\citeauthoryear{Bipat, Bos, Vaish, and
  Monroy-Hern{\'a}ndez}{Bipat et~al\mbox{.}}{2019}]%
        {bipat2019analyzing}
\bibfield{author}{\bibinfo{person}{Taryn Bipat},
  \bibinfo{person}{Maarten~Willem Bos}, \bibinfo{person}{Rajan Vaish}, {and}
  \bibinfo{person}{Andr{\'e}s Monroy-Hern{\'a}ndez}.}
  \bibinfo{year}{2019}\natexlab{}.
\newblock \showarticletitle{Analyzing the use of camera glasses in the wild}.
  In \bibinfo{booktitle}{\emph{Proceedings of the 2019 CHI Conference on Human
  Factors in Computing Systems}}. \bibinfo{pages}{1--8}.
\newblock


\bibitem[\protect\citeauthoryear{Bulling, Blanke, and Schiele}{Bulling
  et~al\mbox{.}}{2014}]%
        {bulling2014tutorial}
\bibfield{author}{\bibinfo{person}{Andreas Bulling}, \bibinfo{person}{Ulf
  Blanke}, {and} \bibinfo{person}{Bernt Schiele}.}
  \bibinfo{year}{2014}\natexlab{}.
\newblock \showarticletitle{A tutorial on human activity recognition using
  body-worn inertial sensors}.
\newblock \bibinfo{journal}{\emph{ACM Computing Surveys (CSUR)}}
  \bibinfo{volume}{46}, \bibinfo{number}{3} (\bibinfo{year}{2014}),
  \bibinfo{pages}{1--33}.
\newblock


\bibitem[\protect\citeauthoryear{Castle, Klein, and Murray}{Castle
  et~al\mbox{.}}{2010}]%
        {castle2010combining}
\bibfield{author}{\bibinfo{person}{Robert~Oliver Castle},
  \bibinfo{person}{Georg Klein}, {and} \bibinfo{person}{David~W Murray}.}
  \bibinfo{year}{2010}\natexlab{}.
\newblock \showarticletitle{Combining monoSLAM with object recognition for
  scene augmentation using a wearable camera}.
\newblock \bibinfo{journal}{\emph{Image and Vision Computing}}
  \bibinfo{volume}{28}, \bibinfo{number}{11} (\bibinfo{year}{2010}),
  \bibinfo{pages}{1548--1556}.
\newblock


\bibitem[\protect\citeauthoryear{Cheng and Robinson}{Cheng and
  Robinson}{2001}]%
        {cheng2001personal}
\bibfield{author}{\bibinfo{person}{Li-Te Cheng} {and} \bibinfo{person}{John
  Robinson}.} \bibinfo{year}{2001}\natexlab{}.
\newblock \showarticletitle{Personal contextual awareness through visual
  focus}.
\newblock \bibinfo{journal}{\emph{IEEE Intelligent systems}}
  \bibinfo{volume}{16}, \bibinfo{number}{3} (\bibinfo{year}{2001}),
  \bibinfo{pages}{16--20}.
\newblock


\bibitem[\protect\citeauthoryear{Chun, Bhattacharya, and Thomaz}{Chun
  et~al\mbox{.}}{2018}]%
        {chun2018detecting}
\bibfield{author}{\bibinfo{person}{Keum~San Chun}, \bibinfo{person}{Sarnab
  Bhattacharya}, {and} \bibinfo{person}{Edison Thomaz}.}
  \bibinfo{year}{2018}\natexlab{}.
\newblock \showarticletitle{Detecting eating episodes by tracking jawbone
  movements with a non-contact wearable sensor}.
\newblock \bibinfo{journal}{\emph{Proceedings of the ACM on Interactive,
  Mobile, Wearable and Ubiquitous Technologies}} \bibinfo{volume}{2},
  \bibinfo{number}{1} (\bibinfo{year}{2018}), \bibinfo{pages}{1--21}.
\newblock


\bibitem[\protect\citeauthoryear{Echterhoff, Haladjian, and
  Br{\"u}gge}{Echterhoff et~al\mbox{.}}{2018a}]%
        {echterhoff2018gait}
\bibfield{author}{\bibinfo{person}{Jessica~Maria Echterhoff},
  \bibinfo{person}{Juan Haladjian}, {and} \bibinfo{person}{Bernd Br{\"u}gge}.}
  \bibinfo{year}{2018}\natexlab{a}.
\newblock \showarticletitle{Gait analysis in horse sports}. In
  \bibinfo{booktitle}{\emph{Proceedings of the Fifth International Conference
  on Animal-Computer Interaction}}. \bibinfo{pages}{1--6}.
\newblock


\bibitem[\protect\citeauthoryear{Echterhoff, Haladjian, and
  Br{\"u}gge}{Echterhoff et~al\mbox{.}}{2018b}]%
        {echterhoff2018gait2}
\bibfield{author}{\bibinfo{person}{Jessica~Maria Echterhoff},
  \bibinfo{person}{Juan Haladjian}, {and} \bibinfo{person}{Bernd Br{\"u}gge}.}
  \bibinfo{year}{2018}\natexlab{b}.
\newblock \showarticletitle{Gait and jump classification in modern equestrian
  sports}. In \bibinfo{booktitle}{\emph{Proceedings of the 2018 ACM
  International Symposium on Wearable Computers}}. \bibinfo{pages}{88--91}.
\newblock


\bibitem[\protect\citeauthoryear{Fiore, Fehr, Bodor, Drenner, Somasundaram, and
  Papanikolopoulos}{Fiore et~al\mbox{.}}{2008}]%
        {fiore2008multi}
\bibfield{author}{\bibinfo{person}{Loren Fiore}, \bibinfo{person}{Duc Fehr},
  \bibinfo{person}{Robot Bodor}, \bibinfo{person}{Andrew Drenner},
  \bibinfo{person}{Guruprasad Somasundaram}, {and} \bibinfo{person}{Nikolaos
  Papanikolopoulos}.} \bibinfo{year}{2008}\natexlab{}.
\newblock \showarticletitle{Multi-camera human activity monitoring}.
\newblock \bibinfo{journal}{\emph{Journal of Intelligent and Robotic Systems}}
  \bibinfo{volume}{52}, \bibinfo{number}{1} (\bibinfo{year}{2008}),
  \bibinfo{pages}{5--43}.
\newblock


\bibitem[\protect\citeauthoryear{Fischer, Greenhalgh, and Benford}{Fischer
  et~al\mbox{.}}{2011}]%
        {fischer2011investigating}
\bibfield{author}{\bibinfo{person}{Joel~E Fischer}, \bibinfo{person}{Chris
  Greenhalgh}, {and} \bibinfo{person}{Steve Benford}.}
  \bibinfo{year}{2011}\natexlab{}.
\newblock \showarticletitle{Investigating episodes of mobile phone activity as
  indicators of opportune moments to deliver notifications}. In
  \bibinfo{booktitle}{\emph{Proceedings of the 13th international conference on
  human computer interaction with mobile devices and services}}.
  \bibinfo{pages}{181--190}.
\newblock


\bibitem[\protect\citeauthoryear{Furnari, Farinella, and Battiato}{Furnari
  et~al\mbox{.}}{2015}]%
        {furnari2015recognizing}
\bibfield{author}{\bibinfo{person}{Antonino Furnari},
  \bibinfo{person}{Giovanni~M Farinella}, {and} \bibinfo{person}{Sebastiano
  Battiato}.} \bibinfo{year}{2015}\natexlab{}.
\newblock \showarticletitle{Recognizing personal contexts from egocentric
  images}. In \bibinfo{booktitle}{\emph{Proceedings of the IEEE International
  Conference on Computer Vision Workshops}}. \bibinfo{pages}{1--9}.
\newblock


\bibitem[\protect\citeauthoryear{Girshick}{Girshick}{2015}]%
        {girshick2015fast}
\bibfield{author}{\bibinfo{person}{Ross Girshick}.}
  \bibinfo{year}{2015}\natexlab{}.
\newblock \showarticletitle{Fast r-cnn}. In
  \bibinfo{booktitle}{\emph{Proceedings of the IEEE international conference on
  computer vision}}. \bibinfo{pages}{1440--1448}.
\newblock


\bibitem[\protect\citeauthoryear{Ha and Choi}{Ha and Choi}{2016}]%
        {ha2016convolutional}
\bibfield{author}{\bibinfo{person}{Sojeong Ha} {and} \bibinfo{person}{Seungjin
  Choi}.} \bibinfo{year}{2016}\natexlab{}.
\newblock \showarticletitle{Convolutional neural networks for human activity
  recognition using multiple accelerometer and gyroscope sensors}. In
  \bibinfo{booktitle}{\emph{2016 international joint conference on neural
  networks (IJCNN)}}. IEEE, \bibinfo{pages}{381--388}.
\newblock


\bibitem[\protect\citeauthoryear{Han, Nandakumar, Philipose, Krishnamurthy, and
  Wetherall}{Han et~al\mbox{.}}{2014}]%
        {han2014glimpsedata}
\bibfield{author}{\bibinfo{person}{Seungyeop Han}, \bibinfo{person}{Rajalakshmi
  Nandakumar}, \bibinfo{person}{Matthai Philipose}, \bibinfo{person}{Arvind
  Krishnamurthy}, {and} \bibinfo{person}{David Wetherall}.}
  \bibinfo{year}{2014}\natexlab{}.
\newblock \showarticletitle{GlimpseData: Towards continuous vision-based
  personal analytics}. In \bibinfo{booktitle}{\emph{Proceedings of the 2014
  workshop on physical analytics}}. \bibinfo{pages}{31--36}.
\newblock


\bibitem[\protect\citeauthoryear{Healey and Picard}{Healey and Picard}{1998}]%
        {healey1998startlecam}
\bibfield{author}{\bibinfo{person}{Jennifer Healey} {and}
  \bibinfo{person}{Rosalind~W Picard}.} \bibinfo{year}{1998}\natexlab{}.
\newblock \showarticletitle{Startlecam: A cybernetic wearable camera}. In
  \bibinfo{booktitle}{\emph{Digest of Papers. Second International Symposium on
  Wearable Computers (Cat. No. 98EX215)}}. IEEE, \bibinfo{pages}{42--49}.
\newblock


\bibitem[\protect\citeauthoryear{Kalantarian, Alshurafa, and
  Sarrafzadeh}{Kalantarian et~al\mbox{.}}{2014}]%
        {kalantarian2014wearable}
\bibfield{author}{\bibinfo{person}{Haik Kalantarian}, \bibinfo{person}{Nabil
  Alshurafa}, {and} \bibinfo{person}{Majid Sarrafzadeh}.}
  \bibinfo{year}{2014}\natexlab{}.
\newblock \showarticletitle{A wearable nutrition monitoring system}. In
  \bibinfo{booktitle}{\emph{2014 11th International Conference on Wearable and
  Implantable Body Sensor Networks}}. IEEE, \bibinfo{pages}{75--80}.
\newblock


\bibitem[\protect\citeauthoryear{Kanade and Hebert}{Kanade and Hebert}{2012}]%
        {kanade2012first}
\bibfield{author}{\bibinfo{person}{Takeo Kanade} {and} \bibinfo{person}{Martial
  Hebert}.} \bibinfo{year}{2012}\natexlab{}.
\newblock \showarticletitle{First-person vision}.
\newblock \bibinfo{journal}{\emph{Proc. IEEE}} \bibinfo{volume}{100},
  \bibinfo{number}{8} (\bibinfo{year}{2012}), \bibinfo{pages}{2442--2453}.
\newblock


\bibitem[\protect\citeauthoryear{Khurana, Ahuja, Yu, Mankoff, Harrison, and
  Goel}{Khurana et~al\mbox{.}}{2018}]%
        {khurana2018gymcam}
\bibfield{author}{\bibinfo{person}{Rushil Khurana}, \bibinfo{person}{Karan
  Ahuja}, \bibinfo{person}{Zac Yu}, \bibinfo{person}{Jennifer Mankoff},
  \bibinfo{person}{Chris Harrison}, {and} \bibinfo{person}{Mayank Goel}.}
  \bibinfo{year}{2018}\natexlab{}.
\newblock \showarticletitle{GymCam: Detecting, recognizing and tracking
  simultaneous exercises in unconstrained scenes}.
\newblock \bibinfo{journal}{\emph{Proceedings of the ACM on Interactive,
  Mobile, Wearable and Ubiquitous Technologies}} \bibinfo{volume}{2},
  \bibinfo{number}{4} (\bibinfo{year}{2018}), \bibinfo{pages}{1--17}.
\newblock


\bibitem[\protect\citeauthoryear{Kitani, Okabe, Sato, and Sugimoto}{Kitani
  et~al\mbox{.}}{2011}]%
        {kitani2011fast}
\bibfield{author}{\bibinfo{person}{Kris~M Kitani}, \bibinfo{person}{Takahiro
  Okabe}, \bibinfo{person}{Yoichi Sato}, {and} \bibinfo{person}{Akihiro
  Sugimoto}.} \bibinfo{year}{2011}\natexlab{}.
\newblock \showarticletitle{Fast unsupervised ego-action learning for
  first-person sports videos}. In \bibinfo{booktitle}{\emph{CVPR 2011}}. IEEE,
  \bibinfo{pages}{3241--3248}.
\newblock


\bibitem[\protect\citeauthoryear{Kong and Tan}{Kong and Tan}{2012}]%
        {kong2012dietcam}
\bibfield{author}{\bibinfo{person}{Fanyu Kong} {and} \bibinfo{person}{Jindong
  Tan}.} \bibinfo{year}{2012}\natexlab{}.
\newblock \showarticletitle{DietCam: Automatic dietary assessment with mobile
  camera phones}.
\newblock \bibinfo{journal}{\emph{Pervasive and Mobile Computing}}
  \bibinfo{volume}{8}, \bibinfo{number}{1} (\bibinfo{year}{2012}),
  \bibinfo{pages}{147--163}.
\newblock


\bibitem[\protect\citeauthoryear{Koufaris, Kambil, and LaBarbera}{Koufaris
  et~al\mbox{.}}{2001}]%
        {koufaris2001consumer}
\bibfield{author}{\bibinfo{person}{Marios Koufaris}, \bibinfo{person}{Ajit
  Kambil}, {and} \bibinfo{person}{Priscilla~Ann LaBarbera}.}
  \bibinfo{year}{2001}\natexlab{}.
\newblock \showarticletitle{Consumer behavior in web-based commerce: an
  empirical study}.
\newblock \bibinfo{journal}{\emph{International journal of electronic
  commerce}} \bibinfo{volume}{6}, \bibinfo{number}{2} (\bibinfo{year}{2001}),
  \bibinfo{pages}{115--138}.
\newblock


\bibitem[\protect\citeauthoryear{Laput and Harrison}{Laput and
  Harrison}{2019a}]%
        {laput2019exploring}
\bibfield{author}{\bibinfo{person}{Gierad Laput} {and} \bibinfo{person}{Chris
  Harrison}.} \bibinfo{year}{2019}\natexlab{a}.
\newblock \showarticletitle{Exploring the Efficacy of Sparse, General-Purpose
  Sensor Constellations for Wide-Area Activity Sensing}.
\newblock \bibinfo{journal}{\emph{Proceedings of the ACM on Interactive,
  Mobile, Wearable and Ubiquitous Technologies}} \bibinfo{volume}{3},
  \bibinfo{number}{2} (\bibinfo{year}{2019}), \bibinfo{pages}{1--19}.
\newblock


\bibitem[\protect\citeauthoryear{Laput and Harrison}{Laput and
  Harrison}{2019b}]%
        {laput2019sensing}
\bibfield{author}{\bibinfo{person}{Gierad Laput} {and} \bibinfo{person}{Chris
  Harrison}.} \bibinfo{year}{2019}\natexlab{b}.
\newblock \showarticletitle{Sensing Fine-Grained Hand Activity with
  Smartwatches}. In \bibinfo{booktitle}{\emph{Proceedings of the 2019 CHI
  Conference on Human Factors in Computing Systems}}. \bibinfo{pages}{1--13}.
\newblock


\bibitem[\protect\citeauthoryear{Lee and Grauman}{Lee and Grauman}{2015}]%
        {lee2015predicting}
\bibfield{author}{\bibinfo{person}{Yong~Jae Lee} {and} \bibinfo{person}{Kristen
  Grauman}.} \bibinfo{year}{2015}\natexlab{}.
\newblock \showarticletitle{Predicting important objects for egocentric video
  summarization}.
\newblock \bibinfo{journal}{\emph{International Journal of Computer Vision}}
  \bibinfo{volume}{114}, \bibinfo{number}{1} (\bibinfo{year}{2015}),
  \bibinfo{pages}{38--55}.
\newblock


\bibitem[\protect\citeauthoryear{Levy and Wolf}{Levy and Wolf}{2015}]%
        {levy2015live}
\bibfield{author}{\bibinfo{person}{Ofir Levy} {and} \bibinfo{person}{Lior
  Wolf}.} \bibinfo{year}{2015}\natexlab{}.
\newblock \showarticletitle{Live repetition counting}. In
  \bibinfo{booktitle}{\emph{Proceedings of the IEEE international conference on
  computer vision}}. \bibinfo{pages}{3020--3028}.
\newblock


\bibitem[\protect\citeauthoryear{Mariakakis, Srinivasan, Rachuri, and
  Mukherji}{Mariakakis et~al\mbox{.}}{2016}]%
        {mariakakis2016watchudrive}
\bibfield{author}{\bibinfo{person}{Alex Mariakakis}, \bibinfo{person}{Vijay
  Srinivasan}, \bibinfo{person}{Kiran Rachuri}, {and} \bibinfo{person}{Abhishek
  Mukherji}.} \bibinfo{year}{2016}\natexlab{}.
\newblock \showarticletitle{Watchudrive: Differentiating drivers and passengers
  using smartwatches}. In \bibinfo{booktitle}{\emph{2016 IEEE International
  Conference on Pervasive Computing and Communication Workshops (PerCom
  Workshops)}}. IEEE, \bibinfo{pages}{1--4}.
\newblock


\bibitem[\protect\citeauthoryear{Mayol, Davison, Tordoff, Molton, and
  Murray}{Mayol et~al\mbox{.}}{2004}]%
        {mayol2004interaction}
\bibfield{author}{\bibinfo{person}{Walterio~W Mayol}, \bibinfo{person}{Andrew~J
  Davison}, \bibinfo{person}{Ben~J Tordoff}, \bibinfo{person}{ND Molton}, {and}
  \bibinfo{person}{David~W Murray}.} \bibinfo{year}{2004}\natexlab{}.
\newblock \showarticletitle{Interaction between hand and wearable camera in 2D
  and 3D environments}. In \bibinfo{booktitle}{\emph{Proc. British Machine
  Vision Conference}}, Vol.~\bibinfo{volume}{2}. \bibinfo{pages}{5}.
\newblock


\bibitem[\protect\citeauthoryear{Mortazavi, Pourhomayoun, Alsheikh, Alshurafa,
  Lee, and Sarrafzadeh}{Mortazavi et~al\mbox{.}}{2014}]%
        {mortazavi2014determining}
\bibfield{author}{\bibinfo{person}{Bobak~Jack Mortazavi},
  \bibinfo{person}{Mohammad Pourhomayoun}, \bibinfo{person}{Gabriel Alsheikh},
  \bibinfo{person}{Nabil Alshurafa}, \bibinfo{person}{Sunghoon~Ivan Lee}, {and}
  \bibinfo{person}{Majid Sarrafzadeh}.} \bibinfo{year}{2014}\natexlab{}.
\newblock \showarticletitle{Determining the single best axis for exercise
  repetition recognition and counting on smartwatches}. In
  \bibinfo{booktitle}{\emph{2014 11th International Conference on Wearable and
  Implantable Body Sensor Networks}}. IEEE, \bibinfo{pages}{33--38}.
\newblock


\bibitem[\protect\citeauthoryear{Newell, Yang, and Deng}{Newell
  et~al\mbox{.}}{2016}]%
        {newell2016stacked}
\bibfield{author}{\bibinfo{person}{Alejandro Newell}, \bibinfo{person}{Kaiyu
  Yang}, {and} \bibinfo{person}{Jia Deng}.} \bibinfo{year}{2016}\natexlab{}.
\newblock \showarticletitle{Stacked hourglass networks for human pose
  estimation}. In \bibinfo{booktitle}{\emph{European conference on computer
  vision}}. Springer, \bibinfo{pages}{483--499}.
\newblock


\bibitem[\protect\citeauthoryear{O'Loughlin, Cullen, McGoldrick, O'Connor,
  Blain, O'Malley, and Warrington}{O'Loughlin et~al\mbox{.}}{2013}]%
        {o2013using}
\bibfield{author}{\bibinfo{person}{Gillian O'Loughlin},
  \bibinfo{person}{Sarah~Jane Cullen}, \bibinfo{person}{Adrian McGoldrick},
  \bibinfo{person}{Siobhan O'Connor}, \bibinfo{person}{Richard Blain},
  \bibinfo{person}{Shane O'Malley}, {and} \bibinfo{person}{Giles~D
  Warrington}.} \bibinfo{year}{2013}\natexlab{}.
\newblock \showarticletitle{Using a wearable camera to increase the accuracy of
  dietary analysis}.
\newblock \bibinfo{journal}{\emph{American journal of preventive medicine}}
  \bibinfo{volume}{44}, \bibinfo{number}{3} (\bibinfo{year}{2013}),
  \bibinfo{pages}{297--301}.
\newblock


\bibitem[\protect\citeauthoryear{Papageorgiou, Oren, and Poggio}{Papageorgiou
  et~al\mbox{.}}{1998}]%
        {papageorgiou1998general}
\bibfield{author}{\bibinfo{person}{Constantine~P Papageorgiou},
  \bibinfo{person}{Michael Oren}, {and} \bibinfo{person}{Tomaso Poggio}.}
  \bibinfo{year}{1998}\natexlab{}.
\newblock \showarticletitle{A general framework for object detection}. In
  \bibinfo{booktitle}{\emph{Sixth International Conference on Computer Vision
  (IEEE Cat. No. 98CH36271)}}. IEEE, \bibinfo{pages}{555--562}.
\newblock


\bibitem[\protect\citeauthoryear{Redmon, Divvala, Girshick, and Farhadi}{Redmon
  et~al\mbox{.}}{2016}]%
        {redmon2016you}
\bibfield{author}{\bibinfo{person}{Joseph Redmon}, \bibinfo{person}{Santosh
  Divvala}, \bibinfo{person}{Ross Girshick}, {and} \bibinfo{person}{Ali
  Farhadi}.} \bibinfo{year}{2016}\natexlab{}.
\newblock \showarticletitle{You only look once: Unified, real-time object
  detection}. In \bibinfo{booktitle}{\emph{Proceedings of the IEEE conference
  on computer vision and pattern recognition}}. \bibinfo{pages}{779--788}.
\newblock


\bibitem[\protect\citeauthoryear{Ren and Philipose}{Ren and Philipose}{2009}]%
        {ren2009egocentric}
\bibfield{author}{\bibinfo{person}{Xiaofeng Ren} {and} \bibinfo{person}{Matthai
  Philipose}.} \bibinfo{year}{2009}\natexlab{}.
\newblock \showarticletitle{Egocentric recognition of handled objects:
  Benchmark and analysis}. In \bibinfo{booktitle}{\emph{2009 IEEE Computer
  Society Conference on Computer Vision and Pattern Recognition Workshops}}.
  IEEE, \bibinfo{pages}{1--8}.
\newblock


\bibitem[\protect\citeauthoryear{{Shroff}, {Smailagic}, and
  {Siewiorek}}{{Shroff} et~al\mbox{.}}{2008}]%
        {4911602}
\bibfield{author}{\bibinfo{person}{G. {Shroff}}, \bibinfo{person}{A.
  {Smailagic}}, {and} \bibinfo{person}{D.~P. {Siewiorek}}.}
  \bibinfo{year}{2008}\natexlab{}.
\newblock \showarticletitle{Wearable context-aware food recognition for calorie
  monitoring}. In \bibinfo{booktitle}{\emph{2008 12th IEEE International
  Symposium on Wearable Computers}}. \bibinfo{pages}{119--120}.
\newblock


\bibitem[\protect\citeauthoryear{Smith, Yin, Feiner, and Nayar}{Smith
  et~al\mbox{.}}{2013}]%
        {smith2013gaze}
\bibfield{author}{\bibinfo{person}{Brian~A Smith}, \bibinfo{person}{Qi Yin},
  \bibinfo{person}{Steven~K Feiner}, {and} \bibinfo{person}{Shree~K Nayar}.}
  \bibinfo{year}{2013}\natexlab{}.
\newblock \showarticletitle{Gaze locking: passive eye contact detection for
  human-object interaction}. In \bibinfo{booktitle}{\emph{Proceedings of the
  26th annual ACM symposium on User interface software and technology}}.
  \bibinfo{pages}{271--280}.
\newblock


\bibitem[\protect\citeauthoryear{Souza, Miyagawa, Melo, and Maciel}{Souza
  et~al\mbox{.}}{2017}]%
        {souza2017wellness}
\bibfield{author}{\bibinfo{person}{Marcos Souza}, \bibinfo{person}{Taynah
  Miyagawa}, \bibinfo{person}{Paulo Melo}, {and} \bibinfo{person}{Francimar
  Maciel}.} \bibinfo{year}{2017}\natexlab{}.
\newblock \showarticletitle{Wellness programs: wearable technologies supporting
  healthy habits and corporate costs reduction}. In
  \bibinfo{booktitle}{\emph{International Conference on Human-Computer
  Interaction}}. Springer, \bibinfo{pages}{293--300}.
\newblock


\bibitem[\protect\citeauthoryear{Starner, Schiele, and Pentland}{Starner
  et~al\mbox{.}}{1998}]%
        {starner1998visual}
\bibfield{author}{\bibinfo{person}{Thad Starner}, \bibinfo{person}{Bernt
  Schiele}, {and} \bibinfo{person}{Alex Pentland}.}
  \bibinfo{year}{1998}\natexlab{}.
\newblock \showarticletitle{Visual contextual awareness in wearable computing}.
  In \bibinfo{booktitle}{\emph{Digest of Papers. Second International Symposium
  on Wearable Computers (Cat. No. 98EX215)}}. IEEE, \bibinfo{pages}{50--57}.
\newblock


\bibitem[\protect\citeauthoryear{Szegedy, Toshev, and Erhan}{Szegedy
  et~al\mbox{.}}{2013}]%
        {szegedy2013deep}
\bibfield{author}{\bibinfo{person}{Christian Szegedy},
  \bibinfo{person}{Alexander Toshev}, {and} \bibinfo{person}{Dumitru Erhan}.}
  \bibinfo{year}{2013}\natexlab{}.
\newblock \showarticletitle{Deep neural networks for object detection}. In
  \bibinfo{booktitle}{\emph{Advances in neural information processing
  systems}}. \bibinfo{pages}{2553--2561}.
\newblock


\bibitem[\protect\citeauthoryear{Tag, Vargo, Gupta, Chernyshov, Kunze, and
  Dingler}{Tag et~al\mbox{.}}{2019}]%
        {tag2019continuous}
\bibfield{author}{\bibinfo{person}{Benjamin Tag}, \bibinfo{person}{Andrew~W
  Vargo}, \bibinfo{person}{Aman Gupta}, \bibinfo{person}{George Chernyshov},
  \bibinfo{person}{Kai Kunze}, {and} \bibinfo{person}{Tilman Dingler}.}
  \bibinfo{year}{2019}\natexlab{}.
\newblock \showarticletitle{Continuous alertness assessments: Using EOG glasses
  to unobtrusively monitor fatigue levels In-The-Wild}. In
  \bibinfo{booktitle}{\emph{Proceedings of the 2019 CHI Conference on Human
  Factors in Computing Systems}}. \bibinfo{pages}{1--12}.
\newblock


\bibitem[\protect\citeauthoryear{Teramoto}{Teramoto}{2018}]%
        {teramoto2018behavioral}
\bibfield{author}{\bibinfo{person}{Wataru Teramoto}.}
  \bibinfo{year}{2018}\natexlab{}.
\newblock \showarticletitle{A behavioral approach to shared mapping of
  peripersonal space between oneself and others}.
\newblock \bibinfo{journal}{\emph{Scientific reports}} \bibinfo{volume}{8},
  \bibinfo{number}{1} (\bibinfo{year}{2018}), \bibinfo{pages}{1--10}.
\newblock


\bibitem[\protect\citeauthoryear{Toshev and Szegedy}{Toshev and
  Szegedy}{2014}]%
        {toshev2014deeppose}
\bibfield{author}{\bibinfo{person}{Alexander Toshev} {and}
  \bibinfo{person}{Christian Szegedy}.} \bibinfo{year}{2014}\natexlab{}.
\newblock \showarticletitle{Deeppose: Human pose estimation via deep neural
  networks}. In \bibinfo{booktitle}{\emph{Proceedings of the IEEE conference on
  computer vision and pattern recognition}}. \bibinfo{pages}{1653--1660}.
\newblock


\bibitem[\protect\citeauthoryear{Vaizman, Ellis, and Lanckriet}{Vaizman
  et~al\mbox{.}}{2017}]%
        {vaizman2017recognizing}
\bibfield{author}{\bibinfo{person}{Yonatan Vaizman}, \bibinfo{person}{Katherine
  Ellis}, {and} \bibinfo{person}{Gert Lanckriet}.}
  \bibinfo{year}{2017}\natexlab{}.
\newblock \showarticletitle{Recognizing detailed human context in the wild from
  smartphones and smartwatches}.
\newblock \bibinfo{journal}{\emph{IEEE Pervasive Computing}}
  \bibinfo{volume}{16}, \bibinfo{number}{4} (\bibinfo{year}{2017}),
  \bibinfo{pages}{62--74}.
\newblock


\bibitem[\protect\citeauthoryear{Vaizman, Weibel, and Lanckriet}{Vaizman
  et~al\mbox{.}}{2018}]%
        {vaizman2018context}
\bibfield{author}{\bibinfo{person}{Yonatan Vaizman}, \bibinfo{person}{Nadir
  Weibel}, {and} \bibinfo{person}{Gert Lanckriet}.}
  \bibinfo{year}{2018}\natexlab{}.
\newblock \showarticletitle{Context recognition in-the-wild: Unified model for
  multi-modal sensors and multi-label classification}.
\newblock \bibinfo{journal}{\emph{Proceedings of the ACM on Interactive,
  Mobile, Wearable and Ubiquitous Technologies}} \bibinfo{volume}{1},
  \bibinfo{number}{4} (\bibinfo{year}{2018}), \bibinfo{pages}{1--22}.
\newblock


\bibitem[\protect\citeauthoryear{Wang, Kogashiwa, Ohta, and Kira}{Wang
  et~al\mbox{.}}{2002}]%
        {wang2002validity}
\bibfield{author}{\bibinfo{person}{Da-Hong Wang}, \bibinfo{person}{Michiko
  Kogashiwa}, \bibinfo{person}{Sachiko Ohta}, {and} \bibinfo{person}{Shohei
  Kira}.} \bibinfo{year}{2002}\natexlab{}.
\newblock \showarticletitle{Validity and reliability of a dietary assessment
  method: the application of a digital camera with a mobile phone card
  attachment}.
\newblock \bibinfo{journal}{\emph{Journal of nutritional science and
  vitaminology}} \bibinfo{volume}{48}, \bibinfo{number}{6}
  (\bibinfo{year}{2002}), \bibinfo{pages}{498--504}.
\newblock


\bibitem[\protect\citeauthoryear{Wang, Lee, Mariakakis, Goel, Gupta, and
  Patel}{Wang et~al\mbox{.}}{2015}]%
        {wang2015magnifisense}
\bibfield{author}{\bibinfo{person}{Edward~J Wang}, \bibinfo{person}{Tien-Jui
  Lee}, \bibinfo{person}{Alex Mariakakis}, \bibinfo{person}{Mayank Goel},
  \bibinfo{person}{Sidhant Gupta}, {and} \bibinfo{person}{Shwetak~N Patel}.}
  \bibinfo{year}{2015}\natexlab{}.
\newblock \showarticletitle{Magnifisense: Inferring device interaction using
  wrist-worn passive magneto-inductive sensors}. In
  \bibinfo{booktitle}{\emph{Proceedings of the 2015 ACM International Joint
  Conference on Pervasive and Ubiquitous Computing}}. \bibinfo{pages}{15--26}.
\newblock


\bibitem[\protect\citeauthoryear{Yamazoe, Utsumi, and Hosaka}{Yamazoe
  et~al\mbox{.}}{2005}]%
        {yamazoe2005body}
\bibfield{author}{\bibinfo{person}{Hirotake Yamazoe}, \bibinfo{person}{Akira
  Utsumi}, {and} \bibinfo{person}{Kenichi Hosaka}.}
  \bibinfo{year}{2005}\natexlab{}.
\newblock \showarticletitle{A body-mounted camera system for capturing
  user-view images without head-mounted camera}. In
  \bibinfo{booktitle}{\emph{Ninth IEEE International Symposium on Wearable
  Computers (ISWC'05)}}. IEEE, \bibinfo{pages}{114--121}.
\newblock


\bibitem[\protect\citeauthoryear{Yamazoe, Utsumi, Hosaka, and Yachida}{Yamazoe
  et~al\mbox{.}}{2007}]%
        {yamazoe2007body}
\bibfield{author}{\bibinfo{person}{Hirotake Yamazoe}, \bibinfo{person}{Akira
  Utsumi}, \bibinfo{person}{Kenichi Hosaka}, {and} \bibinfo{person}{Masahiko
  Yachida}.} \bibinfo{year}{2007}\natexlab{}.
\newblock \showarticletitle{A body-mounted camera system for head-pose
  estimation and user-view image synthesis}.
\newblock \bibinfo{journal}{\emph{Image and Vision Computing}}
  \bibinfo{volume}{25}, \bibinfo{number}{12} (\bibinfo{year}{2007}),
  \bibinfo{pages}{1848--1855}.
\newblock


\bibitem[\protect\citeauthoryear{Ye, Li, Liu, Bridges, Rozga, and Rehg}{Ye
  et~al\mbox{.}}{2015}]%
        {ye2015detecting}
\bibfield{author}{\bibinfo{person}{Zhefan Ye}, \bibinfo{person}{Yin Li},
  \bibinfo{person}{Yun Liu}, \bibinfo{person}{Chanel Bridges},
  \bibinfo{person}{Agata Rozga}, {and} \bibinfo{person}{James~M Rehg}.}
  \bibinfo{year}{2015}\natexlab{}.
\newblock \showarticletitle{Detecting bids for eye contact using a wearable
  camera}. In \bibinfo{booktitle}{\emph{2015 11th IEEE International Conference
  and Workshops on Automatic Face and Gesture Recognition (FG)}},
  Vol.~\bibinfo{volume}{1}. IEEE, \bibinfo{pages}{1--8}.
\newblock


\bibitem[\protect\citeauthoryear{Zhan, Ramos, and Faux}{Zhan
  et~al\mbox{.}}{2012}]%
        {zhan2012activity}
\bibfield{author}{\bibinfo{person}{Kai Zhan}, \bibinfo{person}{Fabio Ramos},
  {and} \bibinfo{person}{Steven Faux}.} \bibinfo{year}{2012}\natexlab{}.
\newblock \showarticletitle{Activity recognition from a wearable camera}. In
  \bibinfo{booktitle}{\emph{2012 12th International Conference on Control
  Automation Robotics \& Vision (ICARCV)}}. IEEE, \bibinfo{pages}{365--370}.
\newblock


\bibitem[\protect\citeauthoryear{Zhang, Bernhart, and Amft}{Zhang
  et~al\mbox{.}}{2016}]%
        {zhang2016diet}
\bibfield{author}{\bibinfo{person}{Rui Zhang}, \bibinfo{person}{Severin
  Bernhart}, {and} \bibinfo{person}{Oliver Amft}.}
  \bibinfo{year}{2016}\natexlab{}.
\newblock \showarticletitle{Diet eyeglasses: Recognising food chewing using EMG
  and smart eyeglasses}. In \bibinfo{booktitle}{\emph{2016 IEEE 13th
  International Conference on Wearable and Implantable Body Sensor Networks
  (BSN)}}. IEEE, \bibinfo{pages}{7--12}.
\newblock


\bibitem[\protect\citeauthoryear{Zhang, Zhao, Nguyen, Xu, Sen, Hester, and
  Alshurafa}{Zhang et~al\mbox{.}}{2019b}]%
        {zhang2019necksense}
\bibfield{author}{\bibinfo{person}{Shibo Zhang}, \bibinfo{person}{Yuqi Zhao},
  \bibinfo{person}{Dzung~Tri Nguyen}, \bibinfo{person}{Runsheng Xu},
  \bibinfo{person}{Sougata Sen}, \bibinfo{person}{Josiah Hester}, {and}
  \bibinfo{person}{Nabil Alshurafa}.} \bibinfo{year}{2019}\natexlab{b}.
\newblock \showarticletitle{NeckSense: A Multi-Sensor Necklace for Detecting
  Eating Activities in Free-Living Conditions}.
\newblock \bibinfo{journal}{\emph{arXiv preprint arXiv:1911.07179}}
  (\bibinfo{year}{2019}).
\newblock


\bibitem[\protect\citeauthoryear{Zhang, Iravantchi, Jin, Kumar, and
  Harrison}{Zhang et~al\mbox{.}}{2019a}]%
        {zhang2019sozu}
\bibfield{author}{\bibinfo{person}{Yang Zhang}, \bibinfo{person}{Yasha
  Iravantchi}, \bibinfo{person}{Haojian Jin}, \bibinfo{person}{Swarun Kumar},
  {and} \bibinfo{person}{Chris Harrison}.} \bibinfo{year}{2019}\natexlab{a}.
\newblock \showarticletitle{Sozu: Self-Powered Radio Tags for Building-Scale
  Activity Sensing}. In \bibinfo{booktitle}{\emph{Proceedings of the 32nd
  Annual ACM Symposium on User Interface Software and Technology}}.
  \bibinfo{pages}{973--985}.
\newblock


\bibitem[\protect\citeauthoryear{Zhu, Nayak, and Roy-Chowdhury}{Zhu
  et~al\mbox{.}}{2012}]%
        {zhu2012context}
\bibfield{author}{\bibinfo{person}{Yingying Zhu}, \bibinfo{person}{Nandita~M
  Nayak}, {and} \bibinfo{person}{Amit~K Roy-Chowdhury}.}
  \bibinfo{year}{2012}\natexlab{}.
\newblock \showarticletitle{Context-aware activity recognition and anomaly
  detection in video}.
\newblock \bibinfo{journal}{\emph{IEEE Journal of Selected Topics in Signal
  Processing}} \bibinfo{volume}{7}, \bibinfo{number}{1} (\bibinfo{year}{2012}),
  \bibinfo{pages}{91--101}.
\newblock


\end{thebibliography}
\pagebreak

\begin{appendices}
\section{Grocery Item Level Labeling}
\textbf{\begin{figure}[ht]
  \centering
 \includegraphics[width=\linewidth]{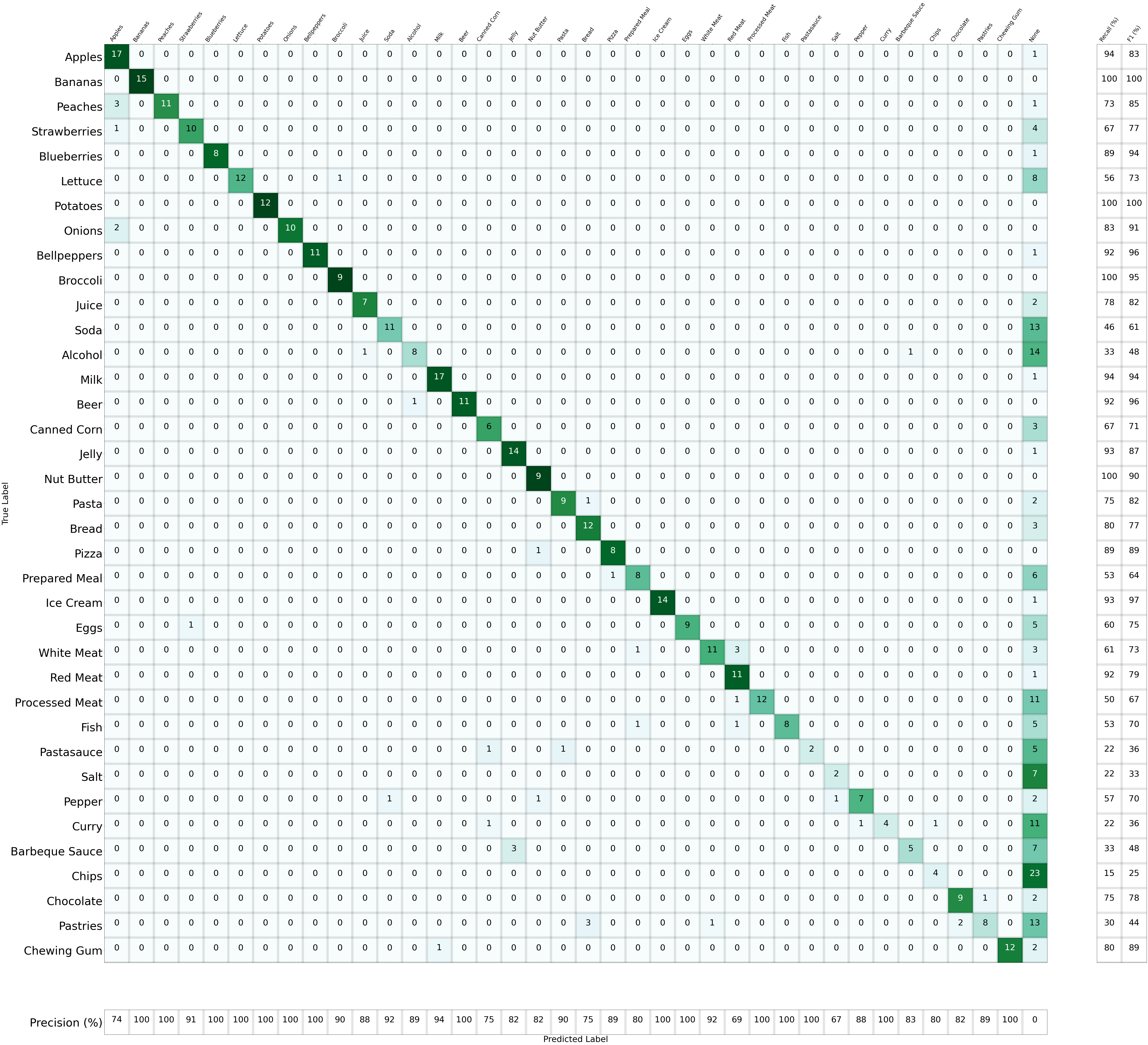}
  \caption{Confusion matrix and precision, recall, f1-score for the shopping item analysis. White boxes indicate zero values.}
  \Description{}
  \label{fig:obj_shop_prf}
\end{figure}}
\end{appendices}

\end{document}